\documentclass[aps,pre,nofootinbib,showpacs,twocolumn,10pt]{revtex4-1}
\usepackage{amssymb,latexsym,mathrsfs}
\usepackage{amsfonts}
\usepackage{mathrsfs}
\usepackage{amsmath}
\usepackage{graphicx}
\usepackage{srcltx}
\usepackage{color}
\usepackage{cancel}
\begin{document}

\title{Strong Fluctuation Theorem for nonstationary Nonequilibrium Systems}
\author{David Luposchainsky$^1$, Andre Cardoso Barato$^2$, and Haye Hinrichsen$^1$}
\affiliation{$^1$Universit\"at W\"urzburg, Fakult\"at f\"ur  Physik und Astronomie, 97074  W\"urzburg, Germany,}
\affiliation{$^2$Universit\"at Stuttgart, II. Institut f\"ur Theoretische Physik, 70550 Stuttgart, Germany.}

\def\d{{\rm d}}
\def\0{\emptyset}
\def\pi{p^{(i)}}
\def\pf{p^{(\infty)}}
\def\at{$@$}

\begin{abstract}
We introduce a finite-time detailed fluctuation theorem of the form $\tilde P(\Delta S_{\rm env}) = e^{\Delta S_{\rm env}} \tilde P(-\Delta S_{\rm env})$ for an appropriately weighted probability density of the external entropy production in the environment. The fluctuation theorem is valid for nonequilibrium systems with constant rates starting with an arbitrary initial probability distribution. We discuss the implication of this new relation for the case of a temperature quench in classical equilibrium systems. The fluctuation theorem is tested numerically for a Markov jump process with six states and for a surface growth model.
\end{abstract}


\email{
	dluposchainsky\at{}physik.uni-wuerzburg.de \\
	barato\at{}theo2.physik.uni-stuttgart.de \\
	hinrichsen\at{}physik.uni-wuerzburg.de
}

\maketitle
\parskip 1mm

\vspace{5mm}

\section{Introduction}

In nonequilibrium statistical physics one of the most important advances in recent years has been the discovery of various fluctuation relations \cite{Evans,Gallavotti,Searles,Kurchan,Lebowitz,Jarzynski,Crooks1,Crooks2,maes99,Sasa,Jia1,Seifert1,andrieux07,harris07,kurchan07,SeifertReview}, which constrain the probability distribution of entropy fluctuations. These relations are of great theoretical importance, being the most general statements for nonequilibrium systems and constituting a generalization of the second law of thermodynamics. Moreover, some of these relations can be tested in real experiments, as for example the Jarzynski relation \cite{Jarzynski}, which relates the work done on a process driven out of equilibrium with equilibrium free energy differences.      

A so-called \textit{detailed fluctuation theorem} (DFT) is a symmetry of the probability distribution $P(\Delta S)$ of some time-integrated quantity $\Delta S$ along the stochastic trajectory of the system. As pointed out by Seifert in a recent review~\cite{SeifertReview}, there are  two types of DFTs. The first one, here denoted as \textit{strong} DFT, is a symmetry of the form
\begin{equation}
\label{SDFT}
P(\Delta S)=e^{\Delta S}P(-\Delta S)
\end{equation}
which relates the positive half with the negative half of the distribution. The other type, called \textit{weak} or \textit{Crooks-type} DFT, is a relation of the form
\begin{equation}
\label{WDFT}
P(\Delta S)=e^{\Delta S}P^\dagger(-\Delta S)
\end{equation}
between two different probability distributions $P$ and $P^\dagger$, where the latter corresponds to some kind of conjugate or reversed process.

DFTs imply identities for averages of certain functions of $\Delta S$, which are known as integral fluctuation theorems (IFTs). It turns out that the two variants differ significantly in their predictive power: A weak DFT implies only a single IFT, namely
\begin{equation}
\label{WIFT}
\langle e^{-\Delta S} \rangle =\int_{-\infty}^{+\infty} \d \Delta S \,\, P(\Delta S)\,\, e^{-\Delta S} = 1\,.
\end{equation}
Using Jensen's inequality for convex functions this IFT induces in turn the second law inequality 
\begin{equation}
\label{SecondLaw}
\langle \Delta S \rangle \geq 0\,.
\end{equation}
Contrarily, a strong DFT implies infinitely many IFTs of the form
\begin{equation}
\label{SIFT}
\langle e^{-\Delta S/2} A(\Delta S) \rangle =\int_{-\infty}^{+\infty}\hspace{-2mm} \d \Delta S \,\, P(\Delta S)\, e^{-\Delta S/2} \,A(\Delta S) =  0,
\end{equation}
where $A(\Delta S)=-A(-\Delta S)$ is an \textit{arbitrary} antisymmetric function. Choosing $A(\Delta S)=\sinh(\Delta S/2)$ one can see that these IFTs include Eq.~(\ref{WIFT}), and therewith the second law~(\ref{SecondLaw}), as a special case.

In this paper we obtain a new strong DFT for Markov jump processes. So far the only quantity which is known to obey a strong DFT is the total entropy production of the system combined with its environment, provided that the transition rates are constant and that the system is stationary throughout the whole observation period. Here we obtain a finite-time strong DFT for constant transition rates, which is valid for relaxation processes, i.e. the initial probability distribution does not need to be the stationary one. Moreover, the quantity entering our DFT is the entropy that flows from the system to the environment  $\Delta S_{\rm env}$~\cite{Seifert1}.

As a concrete application, we consider a temperature quench of an equilibrium system in contact with  single heat bath. We show that the probability distribution of the energy that flows from the reservoir to the system during the relaxation process is constrained by a simple IFT. We also confirm this IFT numerically in a microscopic model for surface growth. Furthermore, we verify the proposed DFT numerically in the case of an explicit nonequilibrium system with six configurations and randomly chosen rates.

The paper is organized as follows: In the next section we present the DFT for the entropy production. As an application we discuss the case of a temperature quench in a equilibrium system in section \ref{sec3}, while section \ref{sec4} contains the numerical tests.  The paper ends with concluding remarks in section~\ref{sec5}. A proof of the DFT can be found in the appendix.

\section{Fluctuation theorem for the entropy production $\Delta S_{\rm env}$}

Let us consider a stochastic Markov process with arbitrary constant rates $w_{c \to c'}$ and an arbitrary initial distribution $\pi_c$. Drawing an initial configuration $c_0$ from this distribution, the process evolves along a certain stochastic path $\gamma$, reaching some final configuration $c_T$ at time~$T$. If $F[\gamma]$ is a functional of the path, its average over many realizations is given by
\begin{equation}
\Bigl\langle F[\gamma] \Bigr\rangle \;=\; \int \mathcal D \gamma \, W[\gamma] \, F[\gamma] \,,
\end{equation}
where the integral runs over all possible stochastic paths~$\gamma$ and $W[\gamma]$ denotes the statistical weight of the path. 
Here we are particularly interested in the entropy production in the external environment $\Delta S_{\rm env}[\gamma]$, which is defined as
\begin{equation}
\Delta S_{\rm env}[\gamma]=  \sum_{j=1}^n \ln \frac{w_{c_{j-1} \to c_j}}{w_{c_{j} \to c_{j-1}}} \;,
\end{equation}
where the path $\gamma$ is such that jumps $c_j \to c_{j+1}$ happen at times $t_j$, $n$ is the total number of jumps along the path, and $c_j$ is the configuration of the system in the time interval $[t_{j},t_{j+1}]$. The corresponding probability distribution can be expressed as
\begin{align}
	P(\Delta S_{\rm env}=X)
	&=
		\Big\langle\delta(X-\Delta S_{\rm env}[\gamma])\Big\rangle \\
	&=
		\int \mathcal D \gamma \, W[\gamma] \,\, \delta\bigl(X -\Delta S_{\rm env}[\gamma]\bigr)\,. \notag
\end{align}
In order to establish a fluctuation theorem for $\Delta S_{\rm env}$, we introduce a \textit{weighted} average
\begin{equation}
\label{WeightedAverage}
\Big\langle F[\gamma] \Big\rangle_{\hspace{-1mm} \chi} \;=\; \frac{1}{\mathcal N} \int \mathcal D \gamma \, W[\gamma] \, F[\gamma] \,\chi_{c_0,c_T} 
\end{equation}
with the additional boundary weights $\chi_{c_0,c_T}$ and a corresponding normalization factor $\mathcal N = \int \mathcal D \gamma \, W[\gamma] \,\chi_{c_0,c_T}$. In the following we shall assume that the boundary weights~$\chi_{c_0,c_T}$ are constrained by the symmetry 
\begin{equation}
\label{Symmetry}
\chi_{c_0,c_T} \, \pi_{c_0} \;=\; \chi_{c_T,c_0}\, \pi_{c_T} \qquad \forall c_0,c_T.
\end{equation}
The corresponding weighted probability density reads
\begin{align}
\label{WeightedProbability}
\tilde P(\Delta S_{\rm env}=X)&=\Big\langle\delta(X-\Delta S_{\rm env}[\gamma])\Big\rangle_{\hspace{-1mm} \chi}\\
&=\frac{1}{\mathcal N}\int \mathcal D \gamma \, W[\gamma] \,\chi_{c_0,c_T} \, \delta\bigl(X -\Delta S[\gamma]_{\rm env}\bigr)\notag\,.
\end{align}
For example, if we choose $\chi_{c_0,c_T} =\pi_{c_T}$, this gives the probability of $\Delta S_{\rm env}$ in an ensemble where each stochastic trajectory is weighted with the initial probability distribution of the final state.

As our main result, we find that this weighted probability density of the external entropy obeys a strong DFT of the form
\begin{align}
\label{thingie}
\tilde P(\Delta S_{\rm env}=X) \;=\; e^X \tilde P(\Delta S_{\rm env}=-X).
\end{align}
A proof of this relation is given in the appendix. This strong DFT implies the IFT
\begin{equation}
\label{antisymmetric}
\left\langle e^{-\Delta S_{\text{env}}/2}\,A(\Delta S_{\text{env}})\right\rangle_{\hspace{-1mm} \chi} = 0
\end{equation}
for arbitrary antisymmetric functions $A(x)$, including the special case
\begin{equation}
\label{NewSecondLaw}
\Big\langle e^{-\Delta S_{\text{env}}} \Big\rangle_{\hspace{-1mm} \chi} = 1 \quad
\Rightarrow \quad \Big\langle \Delta S_{\text{env}} \Big\rangle_{\hspace{-1mm} \chi} \geq 0\,.
\end{equation}
Let us again emphasize that these fluctuation relations are valid for any time $T$ and that the initial probability distribution can be chosen freely. This means that they can be used to study the relaxation into a (equilibrium or nonequilibrium) steady state.

In order to carry out the averages weighted by $\chi_{c_0,c_T}$, it is necessary to know the initial and final configurations of each stochastic trajectory. This information is easily accessible in numerical simulations. In experiments, however, the initial and final configurations are usually not known explicitly. This means that the fluctuation theorem can only be applied if we define the weights in such a way that they can be computed from another experimentally measurable quantity. As an example, we will consider the relaxation of a system with equilibrium dynamics in the following section.

We note that it would be possible to absorb the weights $\chi_{c_0,c_T}$ into the exponential, so that the results above can be written as ordinary (unweighted) averages. However, we think that in general there is not much benefit from such a notation, since as a consequence of this we no longer have a DFT for $\Delta S_{\text{env}}$.

\section{Energy fluctuations after a temperature quench}
\label{sec3}
Let us now restrict to classical equilibrium systems, where each configuration $c \in \Omega$ is associated with a certain internal energy $E_c$. Suppose that the system is initially in thermal equilibrium, in contact with a single heat bath of constant temperature $T_1$, as described by a stationary Boltzmann-Gibbs distribution 
\begin{equation}
\label{StateBefore}
\pi_c = \frac1{Z(\beta_1)} e^{-\beta_1 E_c}\,,
\end{equation}
where $\beta_1=1/T_1$ and $Z(\beta_1)=\sum_ce^{-\beta_1 E_c}$ denotes the partition sum. 

Then, at time $t=0$, let us suddenly increase or decrease the temperature of the heat bath to a different constant value $\beta_2 = 1/T_2$. After the quench the system is expected to relax into a new stationary equilibrium state. Let us select a certain instance of time $T>0$ before this equilibrium is reached and ask the question how the total energy that flows from the reservoir into the system $\Delta E = E_{c_T}-E_{c_0}$ during the relaxation process is distributed.

To answer this question, we first note that the temperature quench does not change the energy functional $E_c$, rather it causes a discontinuous change of the transition rates $w_{c \to c'}$. Before the quench (for $t<0$) the rates are constant and obey detailed balance. At $t=0$ the rates assume a different set of constant values, obeying detailed balance with the new temperature $T_2$, i.e.
\begin{equation}
\frac{w_{c \to c'}}{w_{c' \to c}} \;=\; e^{-\beta_2 (E_{c'}-E_c)} \;=\; e^{-\beta_2 \Delta E_{c \to c'}}\,.
\end{equation}
This means that the entropy that flows from the system to the heat bath in the jump $c \to c'$ is given by
\begin{equation}
\Delta S_{\rm env}^{c \to c'} \;=\; \ln \frac  {w_{c \to c'}}{w_{c' \to c}} \;=\; -\beta_2 \Delta E_{c \to c'}\,.
\end{equation}
Knowing that the rates are constant after the quench, this tells us that the entropy production along a stochastic path $\gamma$ is related to the system energy change by
\begin{equation}
	\Delta S_{\rm env}[\gamma]
	\;=\;
	-\beta_2 \Delta E[\gamma]= -\beta_2(E_{c_T}-E_{c_0}),
\end{equation}
which holds exactly even if the system has not yet reached the new equilibrium state.

To describe the energy fluctuations in this situation, we now define the weights as 
\begin{equation}
\chi_{c_0,c_T} \;\equiv\;\sqrt{{\pi_{c_T}}/{\pi_{c_0}}}\,,
\end{equation}
which obviously fulfills the symmetry condition~(\ref{Symmetry}). With this choice and the given initial state before the quench (\ref{StateBefore}), these weights can be expressed as
\begin{equation}
\label{ChiSpecial}
\chi_{c_0,c_T} \;=\;  e^{-\beta_1 \Delta E/2}\,.
\end{equation}
Inserting this expression into Eq.~(\ref{antisymmetric}) we obtain
\begin{align}
0 &= \left\langle e^{-\Delta S_{\text{env}}/2}\,A(\Delta S_{\text{env}})\right\rangle_{\hspace{-1mm} \chi} \notag \\
&= \left\langle e^{+\beta_2\Delta E/2}\,A(-\beta_2\Delta E)\right\rangle_{\hspace{-1mm} \chi}  \\ \notag
&= \left\langle e^{+\beta_2\Delta E/2}\,A(-\beta_2\Delta E) \frac{\chi_{c_0,c_T}}{\mathcal N} \right\rangle,
\end{align}
where we used Eq.~(\ref{WeightedAverage}) in the last equality. Inserting the weights (\ref{ChiSpecial}) and absorbing the prefactors $-\beta_2$ and \(\mathcal N\) into $\tilde A(x)\equiv A(-\beta_2 x )/\mathcal N$, we obtain the IFT
\begin{equation}
\left\langle e^{\frac12 \Delta \beta \Delta E}\,\tilde A(\Delta E)  \right\rangle = 0,
\end{equation}
where $\Delta\beta = \beta_2-\beta_1$ and $\tilde A$ is an arbitrary antisymmetric function. In this expression the brackets $\langle \ldots \rangle$ denote the \textit{ordinary} (non-weighted) average over many trajectories. With the special choice $\tilde A(x)=\sinh(\frac12 \Delta\beta \,x )$, this expression reduces further to
\begin{equation}
	\left\langle e^{\Delta\beta\,\Delta E} \right\rangle \;=\; 1\,.
	\label{thingie-quench}
\end{equation}
Note that the corresponding 'second law' $\langle  \Delta \beta \Delta E \rangle \leq 0$ simply means that the average energy increases (decreases) if the system is heated up (cooled down) during the temperature quench. We note that a fluctuation relation similar to (\ref{thingie-quench}) has been obtained in \cite{jarz04}.

\section{Numerical tests}
\label{sec4}

\subsection{Height fluctuations in a growth process}
\label{wetting-sim}
\begin{figure}
	\centering
	\includegraphics[width=0.9\linewidth]{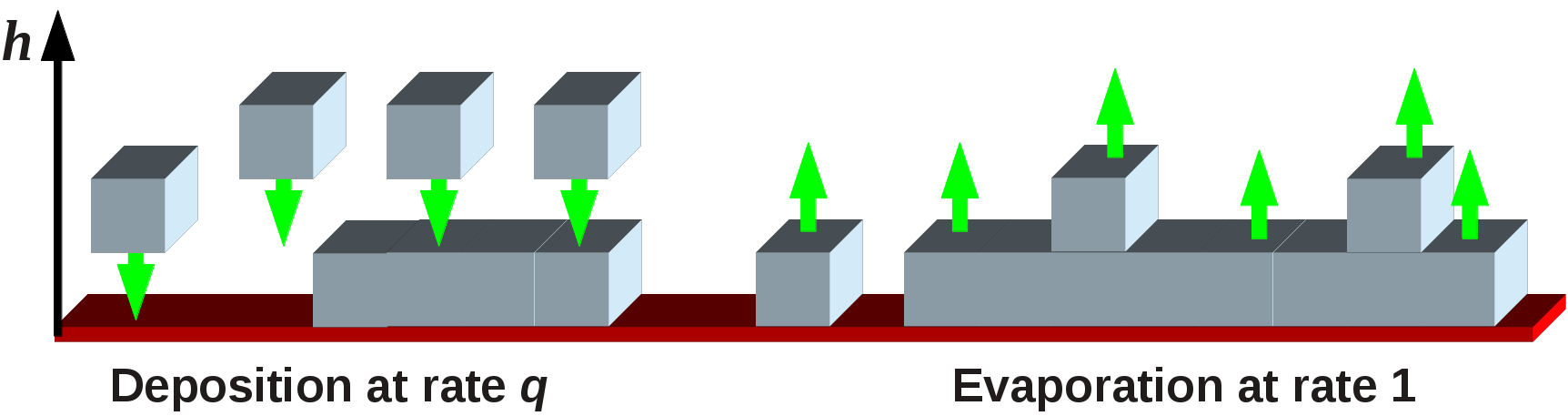}
	\caption[]{(Color online) Growth process on a substrate. Particles are deposited and removed at rate $q$ and $1$, respectively, provided that the resulting configuration does not violate the constraint $|h_i - h_{i \pm 1}| \leq 1$. The green arrows indicate examples of positions where deposition and evaporation is allowed.}
	\label{fig:model}
\end{figure}

To demonstrate the findings of the preceding section, we first consider a solid-on solid growth process on top of an inert substrate, which was investigated some time ago in the context of wetting phenomena~\cite{growth,wetting}. The model is defined on a $d$-dimensional square lattice with periodic boundary conditions, where each site $i$ is associated with the height $h_i=0,1,2,\ldots$ of an interface. It evolves random-sequentially by randomly depositing and removing particles with certain rates. These dynamical rules are constrained by the restriction that neighboring height must not differ by more than one unit, introducing an effective surface tension.

For simplicity we consider here the case of a one-dimensional ring  with \(N\) sites (see Fig.~\ref{fig:model}) with random deposition at rate $q_1$ and evaporation at rate \(1\), subject to the  constraint
\begin{equation}
	| h_i - h_{i+1} | \leq 1 \qquad (h_{N+1} \equiv h_1) ~.
\end{equation}
For $q_1<1$ the system is known to be in a bound state with a stationary probability distribution
\begin{equation}
	P(\{h_i\}) \;\propto\; q_1^H = e^{-\mu H} \,,
\end{equation}
where $H=\sum_{i=1}^N h_i$ is the total number of deposited particles and $\mu=-\ln q$ is the chemical potential. Obviously this is an equilibrium state, where $\mu$ and $H$ play the role of the inverse temperature and the internal energy, respectively. 

\begin{figure}
\includegraphics[width=1.0\linewidth]{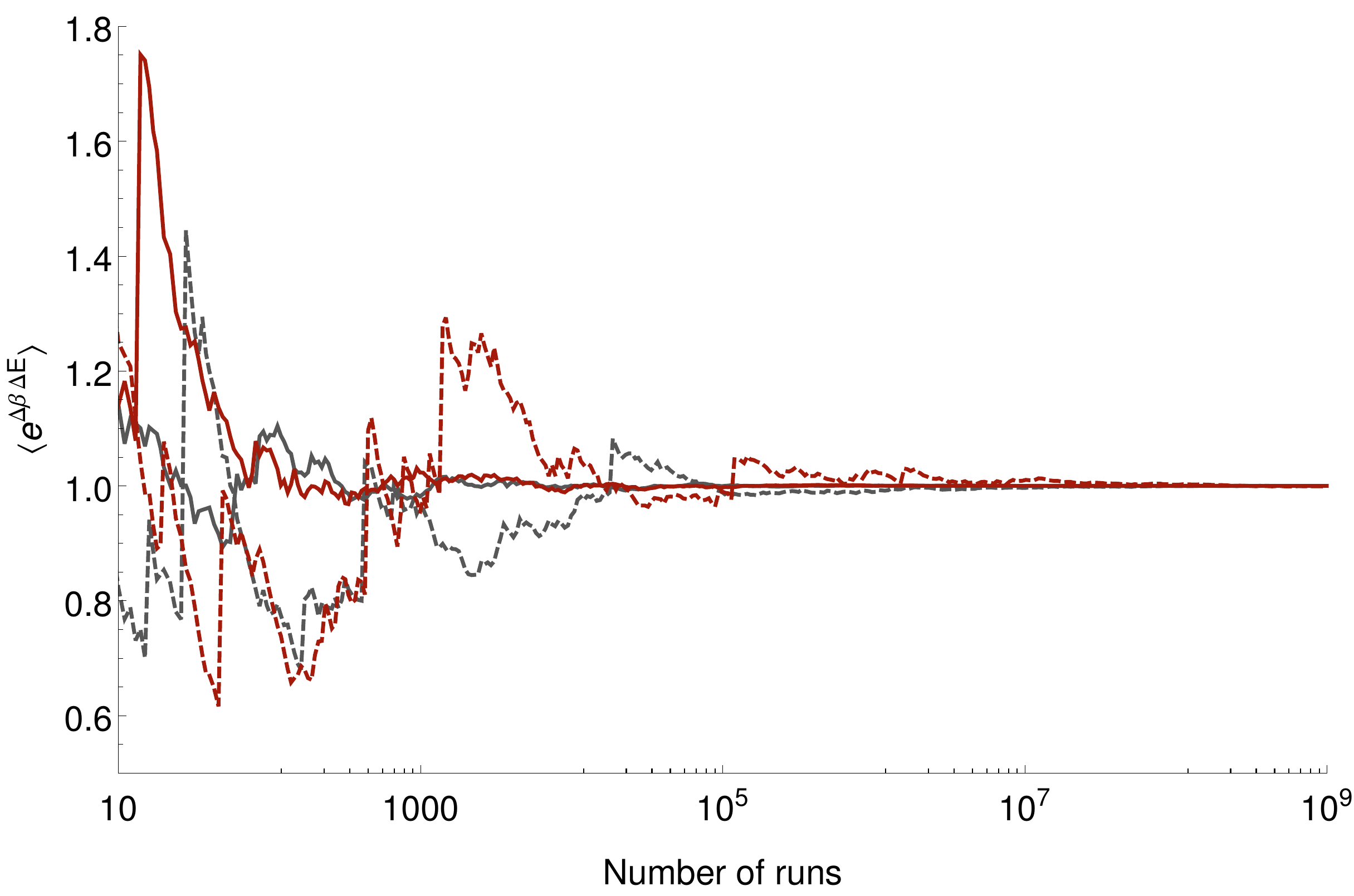}
\caption[]{(Color online)  Convergence of \(\left\langle e^{\Delta\beta\,\Delta E} \right\rangle\) to \(1\) in the growth model with a small system size. Each run starts with an empty lattice and evolves for $50$  time steps to equilibrate. After the quench the process is simulated over $3$ further time steps to reach the final state. Black and red lines denote system sizes \(N=2,4\) respectively; solid lines stand for a quench \(q_1=0.5\rightarrow q_2=0.8\), dotted ones for \(q_1=0.2\rightarrow q_2=0.9\).}
\label{fig:convergence-plot}
\end{figure}

In this stationary state, let us suddenly change the growth rate to a new constant value $q_2<1$ at $t=0$. Subsequently the system relaxes into a new equilibrium state. Applying the results of the preceding section, the fluctuations of the total number of deposited atoms $\Delta H$ between $t=0$ and $t=T$ obey the IFT
\begin{equation}
\left\langle e^{\Delta \mu \Delta H} \right\rangle = 1\,,
\end{equation}
where $\Delta \mu=-(\ln q_2-\ln q_1)$.
In Fig.~\ref{fig:convergence-plot} we show how this average converges to $1$ as the number of runs increases, confirming this IFT in the example of the growth model.

\subsection{Nonequilibrium process with a small state space}
\begin{figure}
	\includegraphics[width=1.0\linewidth]{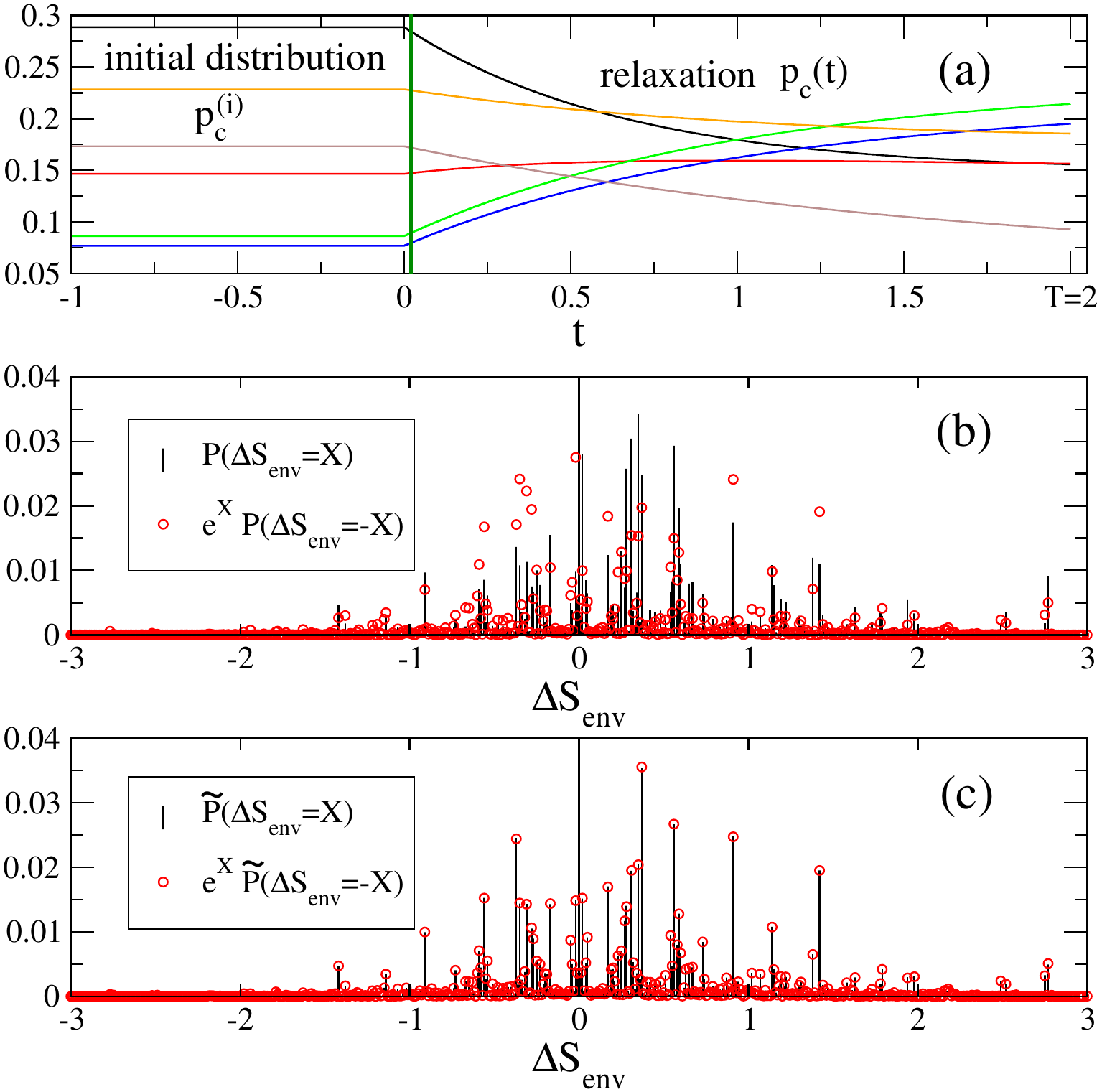}
	\caption[]{(Color online) Simulation of a nonequilibrium process with six configurations. (a) Starting at $t=0$ with a randomly chosen initial distribution $\pi_c$ and randomly chosen constant rates $w_{c\to c'}$ the probability distribution $p_c(t)$ evolves according to the master equation and relaxes into a new nonequilibrium steady state, as shown in the upper panel. (b) Histogram of $P(\Delta S_{\rm env}=X)$ (black bars) together with $e^{X}P(\Delta S_{\rm env}=-X)$ (red dots) taken at $T=2$, demonstrating that the unweighted probability density does not obey a DFT. (c) Contrarily, the weighted probability density $\tilde P(\Delta S_{\rm env}=X)$ defined in Eq.~(\ref{WeightedProbability}) does obey a finite-time DFT, as indicated by the matching of the black bars and the red dots. }
	\label{fig:sixstate}
\end{figure}

The previous examples are special in so far as the initial and the asymptotic final state for $t\to \infty$ obey detailed balance. To demonstrate that our DFT works for any nonequilibrium system with constant rates, we simulated a Markov jump process with 6 configurations and $6^2-6=30$ which are randomly chosen between 0~and~1. Likewise, the initial probability distribution is randomly initialized. Initially the master equation is iterated numerically in order to determine $p_c(T)$, as shown in the upper panel of Fig.~\ref{fig:sixstate}. Then we perform a large number of Monte-Carlo runs, starting with an initial configuration drawn randomly from $\pi_c$ and adding up the contributions to the entropy production whenever the system jumps to a different configuration.

At time $T=2$ the accumulated entropy production is discretized and registered in two histograms. One of them is created as usual by counting the outcomes, whereas the other one is weighted with $\chi_{c_0,c_T}=\pi_{c_T}$. As can be seen in Fig.~\ref{fig:sixstate}b, the data for $P(\Delta S_{\rm env}=X)$ and  $e^{X} P(\Delta S_{\rm env}=-X)$ differ from each other, confirming that the unweighted probability density does not obey a DFT. However, plotting the same data in a weighted histogram, we observe a perfect coincidence in agreement with the DFT (\ref{thingie}) (see Fig.~\ref{fig:sixstate}c).

\section{Conclusions}
\label{sec5}

In this paper we have introduced a strong DFT for the entropy production $\Delta S_{\rm env}$ which is valid for constant rates. In contrast to the usual strong DFT for the total entropy  $\Delta S_{\rm tot}$, which requires the initial state to be stationary, the initial probability distribution can be arbitrary in our case. This means that our DFT is particularly suitable for the study of relaxation processes from arbitrary initial conditions into equilibrium as well as nonequilibrium steady states.

As shown in the appendix, the proof of this DFT follows the same lines as the proofs of other known fluctuation relations. Our strong DFT is particular in so far as it uses an ensemble of trajectories weighted by an extra boundary term $\chi_{c_0,c_T}$, see equation (\ref{WeightedAverage}).

As an application, we have shown that our results imply an IFT for the energy fluctuations during the relaxation of a system in contact with a single heat bath after a sudden temperature quench. In this example the boundary terms $\chi_{c_0,c_T}$ become proportional to the exponential of the energy difference. It would be interesting to find examples of relaxation to a nonequilibrium stationary state where these boundary terms also acquire a clear physical interpretation. Moreover, an experimental verification of relation (\ref{thingie-quench}) should be possible.

It is worth noting that the IFT (\ref{thingie-quench}) for the growth process discussed in section \ref{wetting-sim} seems to hold even if the model quenched from the bound ($q_1<1$) to the moving phase (\(q_2>1\)). We therefore hope that our results may be useful to describe properties of phase transitions, although at this point this remains speculative.

\begin{acknowledgments}
We thank U. Seifert for helpful discussions.
\end{acknowledgments}

\appendix
\section{Proof of the fluctuation theorem}
%
\noindent
Here we present a proof of Eq.~(\ref{thingie}) following the proof of the master FT in \cite{SeifertReview} (see also \cite{garcia}). Let us consider a specific stochastic path $\gamma$ of $n$ spontaneous transitions taking place at times $t_i \in [0,T]$ and the corresponding reversed path $\gamma^\dagger$:
\begin{align*}
\gamma: \ \quad & c_0 \stackrel{t_1}\longrightarrow c_1 \stackrel{t_2}\longrightarrow c_2  \stackrel{t_3}\longrightarrow \ldots \stackrel{t_n}\longrightarrow c_n\\
\gamma^\dagger: \quad & c_n \stackrel{T-t_n}\longrightarrow c_{n-1} \stackrel{T-t_{n-1}}\longrightarrow c_{n-2}  \stackrel{T-t_{n-2}}\longrightarrow \ldots \stackrel{T-t_1}\longrightarrow c_0\,.
\end{align*}
Let the conditional probability $Q[\gamma]$ be the probability of the path $\gamma$ given that the initial state is $c_0$, i.e.
\begin{equation}
	\label{PQDefinition}
	W[\gamma]=\pi_{c_0} \,Q[\gamma]\,.
\end{equation}
Defining the escape 
rates  $\Lambda_j \equiv \sum_{c'} w_{c_j \to c'}$, this conditional probability distribution is written as 
\begin{equation}
	Q[\gamma] = \Bigl[ \prod_{j=1}^n e^{-(t_j-t_{j-1}) \Lambda_{j-1} } \, w_{c_{j-1} \to c_j}\Bigr] \, e^{-(T-t_n) \Lambda_n }\,.
\end{equation}
In the same way, for the reversed path we have $P[\gamma^\dagger]=p_{c_N} Q[\gamma^\dagger]$ and
\begin{equation}
	Q[\gamma^\dagger]
	=
	e^{-t_1 \Lambda_{0} } \, \Bigl[ \prod_{j=1}^n  w_{c_{j} \to c
_{j-1}} \,\, e^{-(t_{j+1}-t_j)\Lambda_{j} }\Bigr] \,.
\end{equation}
Therefore, the ratio of the conditional probabilities is related to the external entropy production by
\begin{equation}
	\label{QuotientQ}
	\frac{Q[\gamma]}{Q[\gamma^\dagger]}
	\;=\;
	\prod_{j=1}^n \frac{w_{c_{j-1} \to c_j}}{w_{c_{j} \to c_{j-1}}} \;=\; e^{\Delta S_{\rm env}[\gamma]}\,.
\end{equation}
Let us now consider an arbitrary functional $F[\gamma]$ which is antisymmetric under path reversal, i.e.
\begin{equation}
\label{AntisymmetryOfF}
F[\gamma^\dagger]\;=\; -F[\gamma].
\end{equation}
Moreover, let $g(x)$ be an arbitrary function applied to this functional. Using (\ref{PQDefinition}), its weighted average according to Eq.~(\ref{WeightedAverage}) is defined by
\begin{equation}
\label{WeightedAverage2}
\Big\langle g \bigl(F[\gamma]\bigr) \Big\rangle_{\hspace{-1mm} \chi} \;=\; 
\frac{1}{\mathcal N}\int \mathcal D \gamma \,\, \pi_{c_0}\,\,Q[\gamma] \,\,\chi_{c_0,c_T} \,\, g\bigl(F[\gamma]\bigr).
\end{equation}
Summing over all reversed paths, the above formula becomes
\begin{equation}
\Big\langle g \bigl(F[\gamma]\bigr) \Big\rangle_{\hspace{-1mm} \chi} 
\;=\; \frac{1}{\mathcal N}\int \mathcal D \gamma^\dagger \,\,\pi_{c_T} \,\,Q[\gamma^\dagger] \,\, \chi_{c_T,c_0}\,\, g\bigl(F[\gamma^\dagger]\bigr) ~.
\end{equation}
Using (\ref{QuotientQ}) and (\ref{AntisymmetryOfF}), we obtain
\begin{equation}
\Big\langle g \bigl(F[\gamma]\bigr) \Big\rangle_{\hspace{-1mm} \chi} 
=\frac{1}{\mathcal N}\int \mathcal D \gamma \,\pi_{c_T} \,e^{-\Delta S_{\rm env}[\gamma]}\,Q[\gamma] \, \chi_{c_T,c_0} \,g\bigl(-F[\gamma]\bigr). 
\end{equation}
Finally, using the symmetry assumed in~(\ref{Symmetry}), i.e.
\begin{equation}
\chi_{c_0,c_T} \, \pi_{c_0} \;=\; \chi_{c_T,c_0}\, \pi_{c_T},
\end{equation}
we have
\begin{equation}
\Big\langle g \bigl(F[\gamma]\bigr) \Big\rangle_{\hspace{-1mm} \chi} 
=
\frac{1}{\mathcal N}\int \mathcal D \gamma \,\pi_{c_0} \,e^{-\Delta S_{\rm env}[\gamma]}\,Q[\gamma] \, \chi_{c_0,c_T} \,g\bigl(-F[\gamma]\bigr) ~,
\end{equation}
which can be rewritten as
\begin{equation}
\label{MasterFT}
\Big\langle g \bigl(F[\gamma]\bigr) \Big\rangle_{\hspace{-1mm} \chi} \;=\; \Big\langle e^{-\Delta S_{\rm env}[\gamma]} g \bigl(-F[\gamma]\bigr) \Big\rangle_{\hspace{-1mm} \chi}. 
\end{equation}
The fluctuation relations presented in this paper can all be derived from this relation. For example, setting $F[\gamma]= \Delta S_{\rm env}[\gamma]$ and $g(\Delta S_{\rm env}[\gamma])\equiv \delta(X-\Delta S_{\rm env}[\gamma])$ one obtains the strong DFT (\ref{thingie}).


\end{document}